\documentclass[twocolumn]{aastex631}
\usepackage{graphicx}	
\usepackage{amsmath}	
\usepackage{amssymb}	

\shorttitle{Precession for the mode change}
\shortauthors{Tong \& Wang}
\graphicspath{{./}{figures/}}

\begin{document}

\title{Precession for the mode change in a gamma-ray pulsar}

\correspondingauthor{H. Tong (tonghao@gzhu.edu.cn); \\ H. H. Wang (wanghh33@mail.sysu.edu.cn)}

\author[0000-0001-7120-4076]{H. Tong}
\affiliation{School of Physics and Materials Science, Guangzhou University, Guangzhou 510006, China}

\author{H. H. Wang}
\affiliation{School of Physics and Engineering, Henan University
of Science and Technology, Luoyang 471023, China}









\begin{abstract}
PSR J2021+4026 is a gamma-ray pulsar having variations in its spin-down rate and gamma-ray flux. Its variations in timing and emission are correlated, e.g., a larger spin-down rate for a low gamma-ray flux. We show that the mode change in PSR J2021+4026 can be understood in the precession scenario. In the precession model, the inclination angle is modulated due to precession. At the same time, the wobble angle may decay with time. This results in damping of the precession. Combined with magnetospheric torque model and the outer gap model, the damped precession can explain: (1) when the inclination angle is larger, the spin-down rate will be larger, accompanied by a lower gamma-ray flux. (2) The variation amplitude of the gamma-ray flux and spin-down rate is smaller than previous results due to the damping of the precession. The modulation period is becoming shorter due to a smaller wobble angle. In the end, we propose that there are two kinds of modulations in pulsars. Long-term modulations in pulsars may be due to precession. Short-term modulations may be of magnetospheric origin.
\end{abstract}



\section{Introduction}

Pulsars are stable clocks in the universe. At the same time, they also have many variabilities. Nulling and mode changing were known in the early time of pulsar studies (Backer 1970a,b). Intermittent pulsars, pulsars with long nulling timescale up to weeks or years, have larger spin-down rate during the on-state than that during the off-state (Kramer et al. 2006). This may due to the presence of magnetospheric particle outflow in the on-state (Kramer et al. 2006; Li et al. 2014). Later more pulsars were found to have correlated timing and emission variations (Lyne et al. 2010; Shaw et al. 2022). The correlation between timing and emission of pulsars may point to a magnetospheric origin.

The possible mechanism for the variation in pulsar timing and emission is not known at present. Change of magnetospheric current (Kramer et al. 2006) or geometry (Timokhin 2010; Lyne et al. 2013), glitches (Keith et al. 2013), external materials (e.g, fallback disk, Li 2006; asteroids, Brook et al. 2014) are several candidates. Precession is also proposed as one of the modulation mechanism when the modulation timescale is long and quasi-periodic (Stairs et al. 2000).

Possible signals of precession were discussed in pulsars (Stairs et al. 2000), magnetars (Makishima et al. 2014; Desvignes et al. 2024), fast radio bursts (Levin et al. 2020; Tong et al. 2020), and accreting neutron stars (Heyl et al. 2023; Zhao et al. 2024). These studies mainly use the precession to explain the long modulation timescale. Precession will also modulate the magnetospheric geometry, which may explain the variations of polarization position angle (Heyl et al. 2023; Desvignes et al. 2024).

PSR J2021+4026 was discovered through blind frequency search using Fermi large area telescope (Abdo et al. 2009). It is the brightest among those reported. It has a frequency of $3.8$ Hz, and frequency derivative about $-7.8\times 10^{-13} \ \rm Hz \ s^{-1}$ (Abdo et al. 2009). Using these timing parameters, its characteristic age is about $77 \rm \ kyrs$, and characteristic magnetic field about $3.9\times 10^{12} \ \rm G$. Possible association with a supernova remnant indicates a distance about $1.5-2.1 \ \rm kpc$ (Ladouceur \& Pineault 2008; Leahy et al. 2013). PSR J2021+4026 is also observed and studied in the X-rays (Lin et al. 2013; Wang et al. 2018; Rigoselli et al. 2021; Razzano et al. 2023). It is radio quiet despite several radio searches (Trepl et al. 2010; Shaw et al. 2023).

PSR J2021+4026 is the first mode changing gamma-ray pulsar (Allafort et al. 2013). It has (1) repeated variations in both the spin-down rate and gamma-ray flux (Takata et al. 2020; Wang et al. 2024; Fiori et al. 2024). The change in spin-down rate and gamma-ray flux is correlated, e.g., a larger spin-down rate for a low gamma-ray flux. (2) The modulation timescale is rather long, about 6 years. Like the case of intermittent pulsars (Kramer et al. 2006), it enables long term monitoring and classification of different states possible. Due to its variability, PSR J2021+4026 is also a target of the Fermi light curve repository (Abdollahi et al. 2023).

Recent observations of PSR J2021+4026 showed that (Wang et al. 2024; Fiori et al. 2024): (1) its modulation timescale becomes shorter, (2) its modulation amplitude for both spin-down rate and gamma-ray flux becomes smaller. These two aspects may point to an origin of damped precession. In the precession scenario: both the spin-down rate and gamma-ray flux are of magnetospheric origin. The magnetosphere of the neutron star is modulated by precession of the neutron star. And the precession may decay with time.

\section{Model calculations}

\subsection{General picture for correlation between timing and emission of pulsars}

\begin{figure}
  \centering
  \includegraphics[width=0.47\textwidth]{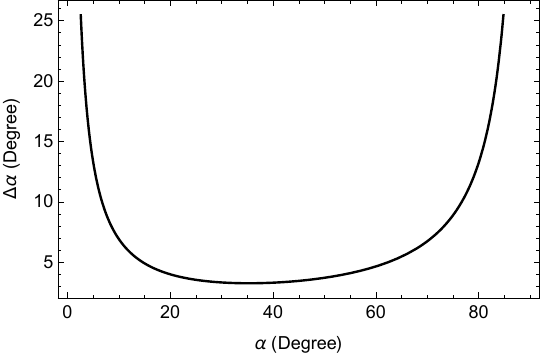}\\
  \caption{Change of the inclination angle required by the change in spin-down torque for PSR J2021+4026.}\label{fig_dalpha}
\end{figure}

\begin{figure}
  \centering
  \includegraphics[width=0.47\textwidth]{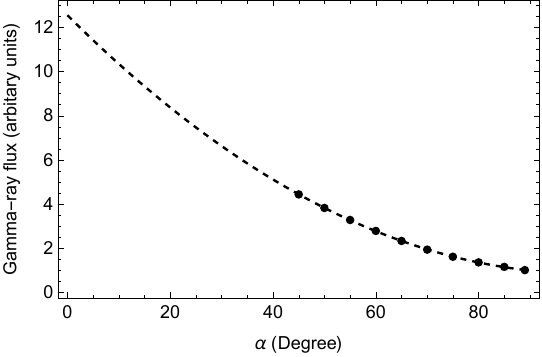}\\
  \caption{Gamma-ray flux above 100 MeV for PSR J2021+4026, as a function of inclination angle. The flux is in arbitrary units. The points are numerical calculations. The dashed line is a quadratic fit of the numerical value: $12.6-13.3 \alpha + 3.8 \alpha^2$ (here $\alpha$ is in units of radians).}\label{fig_gamma_flux_outer_gap}
\end{figure}

The basic observational facts of PSR J2021+4026 is that it has a $4\%$ increase in torque accompanied by a $20\%$ decrease of gamma-ray flux (Allafort et al 2013; Wang et al. 2024; Fiori et al. 2024). In the magnetospheric model of pulsars, the spin-down torque is proportional to: $\propto (1+ \sin^2\alpha)$ (Spitkovsky 2006; Arzamasskiy et al. 2015), where $\alpha$ is the inclination angle (angle between the magnetic axis and the neutron star spin axis). Therefore, a change of the $\alpha$ will result in a change of spin-down torque. For a small change of $\alpha$, the fractional change in the spin-down torque is:
\begin{equation}
  \frac{2\sin\alpha \cos\alpha \Delta\alpha}{1+ \sin^2\alpha} = \frac{\Delta \dot{f}}{\dot{f}}.
\end{equation}
For a fractional change of spin-down torque of $4\%$, the required change in $\alpha$ is shown in figure \ref{fig_dalpha}. From figure \ref{fig_dalpha}, for moderate value of $\alpha$ (20-70 degrees), the required change in $\alpha$ is about $5$ degrees. For $\alpha$ too small or too large, the changes in $\alpha$ is too large to be compatible with the changes in gamma-ray flux (at most $20\%$).

Our calculation for the gamma-ray flux of PSR J2021+4026 is based on the outer gap model. There are many versions of the outer gap model (Romani 1996; Zhang \& Cheng 1997; Zhang et al. 2004; Wang et al. 2010). The self-consistent outer gap (Zhang \& Cheng 1997) is mainly employed here. There are X-ray emissions from PSR J2021+4026 with a surface temperature of $T \sim 10^6 \ \rm K$ (Wang et al. 2018; Rigoselli et al. 2021). These thermal X-ray photons may be the soft X-ray photons in the outer gap closing process. This is a specific case of the general self-consistent outer gap (Zhang \& Cheng 1997; Cheng \& Zhang 2001). In calculating the spectrum of gamma-ray emissions, the particle distribution in the magnetosphere is integrated from a minimum value to a maximum value (eq.(5) in Cheng \& Zhang 2001). Using the local curvature radius to characterise the corresponding radial position, the minimum curvature radius is (in units of light cylinder radius):
\begin{equation}\label{eqn_xmin}
  x_{\rm min} = \frac{2}{3\tan\alpha}.
\end{equation}
The minimum curvature radius (or inner location of the outer gap) depends on the $\alpha$. This will make the gamma-ray flux lower (e.g., flux above 100 MeV) for a higher $\alpha$ (see figure 1 in Cheng \& Zhang 2001, and figure 1 in Tong et al. 2010). Using the parameters of PSR J2021+4026, the gamma-ray flux above 100 MeV as a function of $\alpha$ is shown in figure \ref{fig_gamma_flux_outer_gap}.

From figure \ref{fig_gamma_flux_outer_gap}, it can be seen that when the $\alpha$ is larger, the gamma-ray flux will be lower. Observationally, this means that for a larger torque of the pulsar, the gamma-ray flux will be lower. The timing and emission of PSR J2021+4026 can be understood qualitatively in this way. If the change of $\alpha$ is quasi-periodic, then the variation and correlation of timing and emission will also be in a quasi-periodic way.

Equation (\ref{eqn_xmin}) is only valid for $\alpha> 45$ degree (Cheng \& Zhang 2001). We extrapolate the result to the whole range of $\alpha$ (see figrue \ref{fig_gamma_flux_outer_gap}). Another approximation is: $x_{\rm min} = (2/3) (\pi/2 - \alpha)$ (Romani 1996), which is valid for the whole range of $\alpha$. We have also calculated this case. The general trend for the gamma-ray flux as a function of $\alpha$ is the same. The difference is only quantitative.

\subsection{Quantitative modeling}

\begin{figure}
  \centering
  \includegraphics[width=0.47\textwidth]{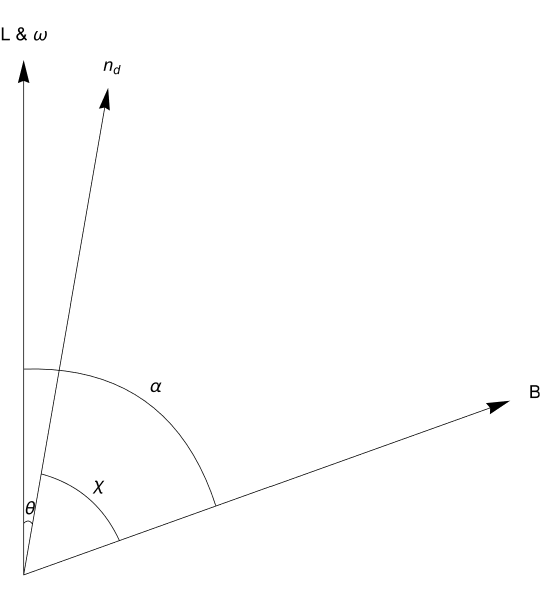}\\
  \caption{Geometry of a precessioning pulsar. The angular momentum $L$ is conserved. The angular velocity $\omega$ is almost coincide with the angular momentum. The angle between the deformation axis $n_d$ and the angular momentum axis is the wobble angle $\theta$. The angle between the magnetic axis and the deformation axis is $\chi$. The deformation axis and the magnetic axis rotates around the angular momentum axis at approximately the pulse frequency $\Omega$. At the same time, the magnetic axis rotates around the deformation axis at the precessional frequency: $\omega_p = \epsilon_0 \Omega \cos\theta$. See also figure 3 in Link \& Epstein (2001),  figure 1 in Jones \& Anderson (2001) and figure 1 in Desvignes et al. (2024).}\label{fig_precession_geometry}
\end{figure}

\begin{figure}
  \centering
  \includegraphics[width=0.47\textwidth]{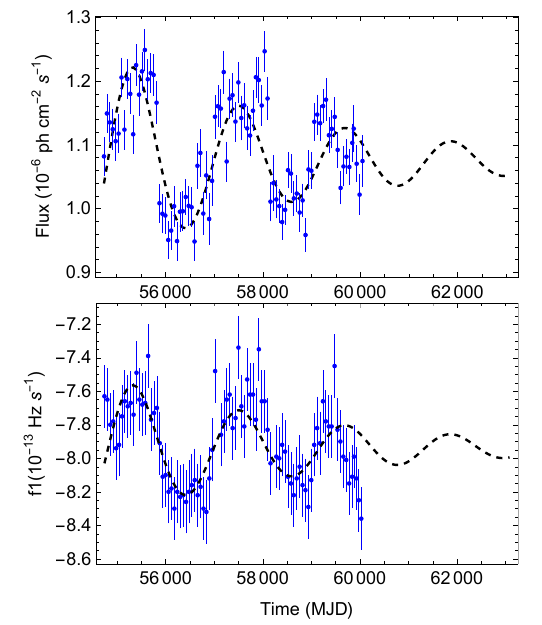}\\
  \caption{Flux and spin-down evolution of PSR J2021+4026. The label ``f1" in the bottom panel denotes the spin-down rate $\dot{f}$. The points and error bars are observational data points from Wang et al. (2024). The dashed lines are the model calculations. }\label{fig_flux_torque_final}
\end{figure}

\begin{figure}
  \centering
  \includegraphics[width=0.47\textwidth]{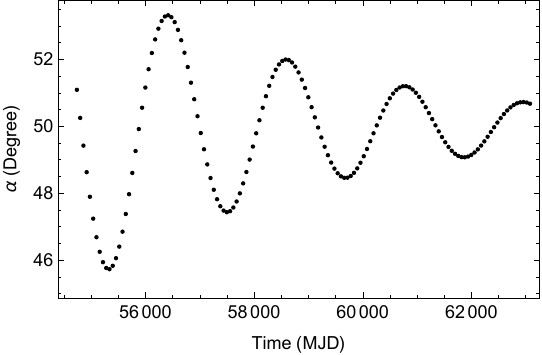}\\
  \caption{Evolution of the inclination angle, expected in the damped precession model. The model parameters are the same as the best fit values in figure \ref{fig_flux_torque_final}.}\label{fig_alpha_evolution}
\end{figure}

The geometry of a precessing pulsar is shown in figure \ref{fig_precession_geometry}.
The general picture is damped precession, for a precessing pulsar with a decaying wobble angle $\theta(t)$ (Jones \& Anderson 2001; Desvignes et al. 2024). In the corotating frame, the angle between the magnetic axis and deformation axis is denoted as $\chi$. From spherical trigonometry or coordinate transformation, the $\alpha$ is:
\begin{equation}\label{eqn_alpha}
  \cos\alpha(t) = \cos\theta(t) \cos\chi + \sin\theta(t) \sin\chi \sin\psi,
\end{equation}
where $\psi$ is the precessional phase. The precessional phase is related to the precessional angular velocity as:
\begin{equation}\label{eqn_rotational_angle}
  \psi = \frac{\pi}{2} - (\omega_p t + \beta_0),
\end{equation}
where $\omega_p$ is the precessional angular velocity, $\beta_0$ is a phase constant. When $\omega_p$ is a function time, the $\omega_p t$ term should replaced by integration.

The precessional angular velocity is related to the pulsar's angular velocity as:
\begin{equation}\label{eqn_precession_angular_velocity}
  \omega_p = \epsilon_0 \Omega \cos\theta(t),
\end{equation}
where $\epsilon_0$ is the ellipticity of the neutron star. Equation (\ref{eqn_rotational_angle}) and (\ref{eqn_precession_angular_velocity}) are textbook results for free precession of an axisymmetric rigid body. Of course, the neutron star structure is more complicated than a rigid body (Jones \& Anderson 2001; Desvignes et al. 2024). These complexities  are neglected. For a small wobble angle, the factor $\cos\theta$ is often omitted (Jones \& Anderson 2001). Due to friction between the crust and core, the wobble angle may decrease with time. The precessional angular velocity will increase with time for a decaying wobble angle. This means that the modulation period will decrease with time for a damped precession.
An exponential decaying form is assumed (Desvignes et al. 2024):
\begin{equation}
  \theta(t) = \theta_0 e^{-t/\tau},
\end{equation}
where $\theta_0$ is the initial wobble angle, and $\tau$ is the decaying timescale of the wobble angle.

For a damped precession, the final $\alpha$ is the $\alpha = \chi$. Therefore, the spin-down rate for a precessing neutron star can be described as:
\begin{equation}\label{eqn_f1}
  \dot{f} = \dot{f}_0 \frac{1+ \sin^2\alpha(t)}{1 + \sin^2\chi},
\end{equation}
where $\dot{f}_0$ is the spin-down rate when the precession is damped away. It is related to the magnetic field and moment of inertia of the neutron star.
There are six parameters in the above equations: $\dot{f}_0$, $\chi$, $\theta_0$, $\tau$, $\epsilon_0$, and $\beta_0$.

The timing and emission of PSR J2021+4026 has been monitored for more than 10 years (Wang et al. 2024; Fiori et al. 2024). The spin-down evolution can be model using equation (\ref{eqn_f1}), shown in figure \ref{fig_flux_torque_final}. For the theoretical curve of the spin-down rate (bottom panel in figure \ref{fig_flux_torque_final}), $\chi$ is fixed to be $50$ degrees. The other five parameters are: $\dot{f}_0 =-7.94 \times 10^{-13} \ \rm Hz \ s^{-1}$, $\theta_0 = 4.9$ degrees, $\tau=11.7 $ years, $\epsilon_0 = 1.4\times 10^{-9}$, and $\beta_0 =-1.76$. These five parameters are obtained by nonlinear model fitting to the observational data of spin-down rate, using the central values only. If $\chi$ is left free, the best fitted value is about $40$ degrees. However, the value of $\chi$ can not be well constrained. As can be seen from figure \ref{fig_dalpha}, for a $\Delta \alpha$ around 5 degrees, the allowed value of $\alpha$ can be in a wide range. Since $\alpha(t)$ varies around $\chi$, this may also render the angle $\chi$ poorly constrained. By fixing $\chi$ from 20 degree to 70 degree, we have calculated the corresponding gamma-ray flux light curves. Combined with the gamma-ray observations, we choose a value of $\chi=50$ degrees as an optimal guess. Therefore, the theoretical results in figure \ref{fig_flux_torque_final} are model calculations for typical parameters. We want to show that there are some points in the parameter space that can simultaneously model the timing and emission of PSR J2021+4026 in the precession scenario.

When the rotational evolution of the pulsar has been modeled (dashed line in the bottom panel of figure \ref{fig_flux_torque_final}), the $\alpha$ can be found as a function of time, shown in figure \ref{fig_alpha_evolution}. Combined with the calculations of the outer gap model (figure \ref{fig_gamma_flux_outer_gap}), the expected gamma-ray flux as a function of time can be obtained. This is the theoretical curve in the upper panel in figure \ref{fig_flux_torque_final}. From figure \ref{fig_flux_torque_final}, both the timing and emission of PSR J2021+4026 can be modeled quantitatively.

From figure \ref{fig_alpha_evolution}, the $\alpha$ of PSR J2021+4026 varies around 50 degrees (the value of $\chi$). The pulse profiles of gamma-ray pulsars can be explained using a combination of inclination angle and viewing angle (Watters et al. 2009; Kalapotharakos et al. 2014). The gamma-ray pulse profile of PSR J2021+4026 may be explained by an $\alpha$ about (40-60) degrees (Trepl et al. 2010). While X-ray modeling of the pulse profile of PSR J2021+4026 found an $\alpha$ about ($20-25$) degrees (Rigoselli et al. 2021). An $\alpha$ about $60$ degrees is employed in modeling    both the X-ray and gamma-ray pulse profiles (Wang et al. 2018). At present, we are not sure about the exact  value of $\alpha$ for PSR J2021+4026.

\section{Discussion}

Since the discovery of variations in PSR J2021+4026 (Allafort et al. 2013; Wang et al. 2024), precession is always proposed as one of the physical origins, along with many other possibilities. We have shown that the general trend of the timing and emission behaviors of PSR J2021+4026 can be modeled quantitatively in the precession scenario. Our main results are presented in figure \ref{fig_flux_torque_final}. A pulsar magnetosphere modulated by a damped precession can explain: (1) the variations and correlations in the torque and gamma-ray flux. (2) the modulation period becomes shorter and the amplitude becomes smaller (Wang et al. 2024). Of course, during the calculations there are many assumptions and uncertainties.
\begin{enumerate}
  \item In modeling the gamma-ray flux, we mainly focus on the relative change of the gamma-ray flux. The absolute value of observed gamma-ray flux depends on the beaming, distance etc (Zhang \& Cheng 1997; Cheng \& Zhang 2001). By choosing the appropriate parameters, the absolute gamma-ray flux may also be explained quantitatively.

  \item In explaining the gamma-ray flux, we mainly focus on the integral flux above 100 MeV. During the mode change, the spectra and pulse profile of PSR J2021+4026 also changes (Allafort et al. 2013; Wang et al. 2024; Fiori et al. 2024). For a larger $\alpha$, the spin-down torque is larger. The gamma-ray flux is expected to be lower. The spectra is expected to be harder (see figure 1 in Cheng \& Zhang 2001, and figure 1 in Tong et al. 2010). There are marginal evidence for a harder spectra during the low gamma-ray flux state (figure 4 in Wang et al. 2024). Explaining the pulse profile change is beyond the scope of self-consistent outer gap (Zhang \& Cheng 1997). Three dimensional modeling of the outer gap is required. Previous studies showed that a change in the $\alpha$ may explain the change in the pulse profile (Ng et al. 2016).

  \item In explaining the spin-down behaviors, the magnetohydrodynamical torque is assumed (proportional to $1+\sin^2 \alpha$). Considering particle acceleration and particle wind outflow, the pulsar spin-down torque may have different forms. For example, the torque may be proportional to $(\kappa + \sin^2\alpha$), where $\kappa$ is the particle contribution to the spin-down torque (Kou \& Tong 2015). By including this additional parameter, the comparison between observations and theoretical calculations may be improved. Future more data are needed to determine whether this degree of freedom should be considered (currently it just has $\sim 2.5$ cycles of precession, see figure \ref{fig_flux_torque_final}).

  \item The earlier observations of PSR J2021+4026 indicated that the state change timescale is shorter than one week (Allafort et al. 2013). Such abrupt changes can not be explained by the precession scenario. It is possible that the modulation is till due to precession of the neutron star, while other physics must be added to explain the abrupt state change. Several speculations, such as existence of threshold for pair production process etc, are discussed in Jones (2012). While, the later state changes of PSR J2021+4026 is more likely to be smooth (Wang et al. 2024; Fiori et al. 2024). This is consistent with the expectation of precession scenario.

  \item The damping of the precession is modeled as a decaying wobble angle (Desvignes et al. 2024). There are also other ways to model a damped precession, e.g., a changing ellipticity etc. The damping mechanism may be due to friction of the crust and core. Physical modeling of a damped precession is required.

  \item Although a damped precession may explain the state change in PSR J2021+4026, the trigger of the precession is unknown at present. A star quake, or plate tectonics may trigger the precession of the neutron star. The star quake/plate tectonics may also trigger a glitch in the neutron star. Of course, the coexistence of precession and glitch is a complicated problem (Jones et al. 2017). If the whole star is in a solid state (e.g., a solid quark star, Xu 2003), the underlying physics will be very different for both precession and glitch.

\end{enumerate}

We mainly focus on the observations of rotation and gamma-ray flux. Only the $\alpha$ is required. For a viewing angle about $90$ degrees (Allafort et al. 2013), PSR J2021+4026 is expected to be radio quiet. The precession mainly provides a clock mechanism for the pulsar magnetosphere. In order to explain the spin-down behavior and gamma-ray flux, a spin-down torque and a gamma-ray emission mechanism are required. Similarly, in order to explain the X-ray pulse profile and its phase shift with the gamma-ray pulse profile, additional physical input may be required, e.g., a multipole magnetic field (Wang et al. 2018; Razzano et al. 2023).

\section{Conclusion}

Since the discovery of nulling and mode changing in pulsars (Backer 1970a,b), there are many physical models proposed (Kramer et al. 2006; Lyne et al. 2010). The modulation period of PSR J2021+4026 is relatively long (about 6 years), thus it provides us an opportunity to study the mechanism of mode change in pulsars. Through our study, we want to propose that:
\begin{enumerate}
  \item All long term mode changes (including nulling) may due to precession of the neutron star. For a nearly spherical neutron star, precession period is a natural long modulation timescale (Jones 2012). The precession may be damped to different degree in different sources (Desvignes et al. 2024). The precession may be triggered by star quake or plate tectonics etc. Not only the mode change of PSR J2021+4026, but also the radio mode changes (Lyne et al. 2010; Shaow et al. 2022) may all due to precession. The key point is: precession mainly provides the clock mechanism. In order to explain the rotational behaviors, a spin-down torque should be employed. In order to explain the emission variations (e.g., gamma-ray flux observations of PSR J2021+4026), an emission mechanism should be employed. The spin-down torque,  emission mechanism, and all other physical process in the magnetosphere are modulated by the precession.

  \item All short term ($\sim 10$ pulsar period) mode changes (including nulling) should be of magnetospheric origin. The typical length scale of the magnetosphere is the light cylinder radius. Therefore, the typical timescale in the magnetosphere will naturally be about several periods. We do not know the exact boundary between long term and short term mode changes at present.

  \item The correlation between radio and high-energy emissions depends strongly the geometry of the neutron star. The gamma-ray flux (which may also be true for the X-ray flux) depends weakly on the geometry (e.g., $\alpha$, see figure \ref{fig_gamma_flux_outer_gap}). While, the radio flux may be very sensitive on the $\alpha$, and viewing angle etc. Therefore, for a pulsar with a magnetosphere modulated by some process, the correlation between radio observations and high energy observations can be either positive or negative (Hermsen et al. 2013; Cao et al. 2024).
\end{enumerate}

The precession model of PSR J2021+4026 have clear predictions: (1) its torque and flux is expected to vary with decreasing amplitude and decreasing period in the future (figure \ref{fig_flux_torque_final}). This may be tested with future monitoring of this source. (2) The $\alpha$ in different state is expected to be different (figure \ref{fig_alpha_evolution}). Since PSR J2021+4026 is radio quiet, future X-ray polarization observations may help to test this point (or radio/X-ray observations of other mode changing pulsars).

\begin{acknowledgments}
This work is supported by National SKA Program of China (No. 2020SKA0120300) and NSFC (12133004).
\end{acknowledgments}









\end{document}